\documentclass{article}
              
            \newcommand{\adj}{a^{\dag}_{\a,j}}
            \newcommand{\adjp}{a^{\dag}_{\a^{\p},j^{\p}}}
            \newcommand{\aj}{a_{\a,j}}
            \newcommand{\ajp}{a_{\a^{\p},j^{\p}}}
            \newcommand{\ad}{a^{\dag}}
            \newcommand{\p}{\prime}

            \newcommand{\ap}{\alpha^{\p}}
            \renewcommand{\sp}{s^{\p}}

            \renewcommand{\a}{\alpha}
            
            \newcommand{\cd}{c^{\dag}}

            \newcommand{\g}{\gamma}
            \newcommand{\gp}{\gamma^{\p}}

            \usepackage{graphics}
            \begin{document}
           \title{Fields of Quantum Reference Frames based on Different
           Representations of Rational Numbers as States of Qubit Strings}
          \author{Paul Benioff\\
           Physics Division, Argonne National Laboratory \\
           Argonne, IL 60439 \\
           e-mail: pbenioff@anl.gov}
           \date{\today}
           \maketitle

          \begin{abstract}In this paper fields of quantum
          reference frames based on gauge transformations of
          rational string states are described in a way that,
          hopefully, makes them more understandable than their
          description in an earlier paper.  The approach taken
          here is based on three main points: (1) There are a
          large number of different quantum theory representations
          of natural numbers, integers, and rational numbers as
          states of qubit strings. (2) For each representation,
          Cauchy sequences of rational string states give a
          representation of the real (and complex) numbers. A
          reference frame is associated to each representation.
          (3) Each frame contains a representation of all
          mathematical and physical theories that have the real
          and complex numbers as a scalar base for the theories.
          These points and other aspects of the resulting fields are
          then discussed and justified in some detail. Also two
          different methods of relating the frame field to physics
          are discussed.
          \end{abstract}

          \section{Introduction}
          In other work \cite{BenFIQRF} two dimensional fields of
          quantum reference frames were described that were based on
          different quantum theory representations of the real
          numbers.  Because the description of the fields does not
          seem to be understood, it is worthwhile to approach a description
          in a way that will, hopefully, help to make the fields
          better understood.  This the goal of this contribution to the
          third Feynman conference proceedings.

          The approach taken here is based on three main points:
          \begin{itemize} \item  There are a large number of
          different quantum theory representations of
          natural numbers, integers, and rational numbers as
          states of qubit strings. These arise from gauge
          transformations of the qubit string states. \item
          For each representation, Cauchy sequences of rational
          string states give a representation of the real (and
          complex) numbers. A reference frame is
          associated to each representation. \item Each frame
          contains a representation of all mathematical and physical
          theories. Each of these is a mathematical structure that
          is based on the real and complex number
          representation base of the frame.  \end{itemize} This approach
          is summarized in the next section with more details given
          in the following sections.

          \section{Summary Discussion of the Three Points}\label{SDTP}
          As is well known, the large amount of work and interest in quantum
          computing and quantum information theory is based on quantum
          mechanical representations of numbers as states of strings of
          qubits and linear superpositions of these states. The
          numbers represented by these states are usually the nonnegative
          integers or the natural numbers.  Examples of integer
          representations are states of the form $|\g,s\rangle$ and
          their linear superpositions $\psi =\sum_{s,\g}c(\g,s)|\g,s\rangle.$
         Here $|\g,s\rangle =|\g,0\rangle\otimes_{j=1}^{j=n}
         |s(j),j\rangle$ is a state of a string
          of $n+1$ qubits where the qubit at location $0$ denotes
          the sign (as $\g =+,-$) and $s:[1,\cdots,n]\rightarrow
          \{0,1\}$ is a $0-1$ valued function on the integer
          interval $[1,n]$ where $s(n)=1.$  This last condition
          removes the redundancy of leading $0s.$

          This description can be extended to give quantum mechanical
          representations of rational numbers as states of qubit strings
          \cite{BenRCRNQM} and of real and complex numbers as Cauchy sequences
          of rational number states of qubit strings \cite{BenRRCNQT}.
          As will be seen, string rational numbers can be
          represented by qubit string states $|\g,s\rangle$ where $s$
          is  a $0-1$ valued function on an integer interval $[l,u]$ with
          $l\leq 0$ and $u\geq 0$ and the sign qubit  $\g$, at position
          $0$.

          A basic point to note is that there are a great many
          different representations of rational numbers (and of
          integers and natural numbers) as states of qubit strings.
          Besides the arbitrariness of the location of the qubit
          string on an integer lattice there is the arbitrariness of
          the choice of quantization axis for each qubit in the
          string. This latter arbitrariness is equivalent to a gauge
          freedom for the choice of which system states correspond to
          the qubit states $|0\rangle$ and $|1\rangle$.

          This arbitrariness of gauge choice for each qubit is
          discussed in Section \ref{GT} in terms of  global and
          local gauge transformations of rational string states of
          qubit strings.  A different representation of string rational
          numbers as states of qubit strings is associated with each
          gauge transformation.

          As will be seen in the next section, there is a quantum
          representation of real numbers associated with each
          representation of the rational string states. It is clear
          that there are a large number of different representations
          of real numbers as each representation is associated with a
          different gauge gauge representation of the rational string states.

          The gauge freedom in the choice of quantization axis for
          each qubit plays an important role in quantum cryptography
          and the transfer of quantum information between a sender
          and receiver. The choice of axis is often referred to
          a reference frame  chosen by sender and receiver for
          transformation and receipt of quantum information
          \cite{Bagan,Rudolph,Bartlett,vanEnk}.

          Here this idea of reference frames is taken over in
          that a reference frame $F_{R_{U}}$ is associated with
          each quantum representation $R_{U}$ of real numbers.
          Since each real number representation is associated
          with a gauge transformation $U$, one can also associate
          frames directly with gauge transformations, as in
          $U\rightarrow F_{U}$ instead of $F_{R_{U}}.$

          It should be noted that complex numbers are also included
          in this description since they can be represented as an
          ordered pair of real numbers. Or they can be built up
          directly from complex rational string states.  From now
          on real numbers and their representations will be assumed
          to also include complex numbers and their representations.

          An important point for this paper is that any physical or
          mathematical theory, for which the real numbers form a base
          or scalar set, has a representation in each frame as a
          mathematical structure based on the real number representation
          in the frame. Since this is the case for all physical theories
          considered to date, it follows that they all have representations
          in each frame as mathematical structures based on the real number
          representation in the frame. It follows that theories such
          as quantum mechanics, quantum field theory. QED, QCD,
          string theory, and special and general relativity all have
          representations in each frame as mathematical structures
          based on the real number representation associated with
          the frame. It is also the case that, if the space time
          manifold is considered to be a $4-tuple$ of the real
          numbers, then each frame contains a representation of
          the space time manifold as a $4-tuple$ of the real
          number representation.

          To understand these observations better it is useful to
          briefly describe theories. Here the usual
          mathematical logic \cite{Shoenfield} characterization of
          theories as being defined by a set of axioms is used. All
          theories have in common the logical axioms  and logical
          rules of deduction; they are distinguished by having
          different sets of nonlogical axioms. This is the case
          whether the axioms are explicitly stated or not.\footnote{
          The importance of the axiomatic
          characterization of physical or mathematical theories is
          to be emphasized.  All properties of physical or mathematical
          systems are described in the theory as theorems which are
          obtained from the axioms by use of the logical rules of
          deduction. Without axioms a theory is empty as it
          can not make any predictions or have any meaning.}

          Each theory described by a consistent set of axioms has
          many representations (called models in mathematical
          logic)\footnote{In physics models have a different meaning
          in that they are also theories. However they are simpler
          theories as they are based on simplifying model assumptions
          which serve as axioms for the simpler theory.}
          as mathematical structures in which the theory axioms are
          true.  Depending on the axiom set the representations may
          or may not be isomorphic.

          The real numbers axiomatized as a complete ordered field
          are an example of a simple theory.  For this
          axiomatization all representations of the real numbers are
          isomorphic.  However they are not the same. All theories
          based on the real (or complex) numbers include
          the real (and complex) number axioms in their axiom
          sets.
          These well known aspects of theories are are quite familiar
          in the application of group theory to physics. Each abstract
          group is defined by a set of axioms that consist of the general
          group axioms and additional ones to describe the particular
          group considered. Each group has many different representations
          as different mathematical systems. These can be matrices or
          operators in quantum theory. These are further classified by
          the dimensionality of the representation and whether they are
          reducible or irreducible. The importance of different
          irreducible representations to describe physical systems and
          their states is well known.

          As sets of axioms and derived theorems, physical theories,
          unlike mathematical theories, also have representations as
          descriptions of physical systems and their properties. It
          is immaterial here whether these representations are
          considered to be maps from theory statements to physical
          systems and their properties or from different mathematical
          representations of the theory to physical systems and their
          properties. It is, however, relevant to note that theoretical
          predictions to be tested by experiment are or can be
          represented by equations whose solutions are real numbers.
          Spectra of excited states of nuclei, atoms, and molecules
          are examples.  The same holds for observables with discrete
          spectra such as spin, isospin, and angular momentum.  Here
          the eigenvalues are the real number equivalents of integers
          or rational numbers.

          \section{Real Numbers and their Representation in Quantum Theory}
          \label{RNRQT}It is useful to begin with a question "What are the
          real numbers?" The most general answer is that
          they are elements of any set of mathematical or physical
          systems that satisfies the real
          number axioms.  The axioms express the requirements
          that the set must be a complete, ordered field.
          This means that the set must be closed under
          addition, multiplication and their inverses, A linear
          order must exist for the collection, and any convergent
          sequence of elements converges to an element of the set.

          It follows that all sets of real numbers must at least
          have these properties.  However they can have other
          properties as well. A study of most any mathematical
          analysis textbook  will show that real numbers are
          defined as equivalence classes of either Dedekind
          cuts or of Cauchy sequences of rational numbers.\footnote{
          Rational numbers are defined as equivalence classes of
          ordered pairs of integers which are in turn defined as
          equivalence classes of ordered pairs of the natural
          numbers $0,1,2,\cdots$.} A sequence $t_{n}$ of rational
          numbers is a Cauchy sequence if\begin{equation}\label{cauchynos}
          \begin{array}{c}\mbox{For all $l$ there is an $h$ such
          that} \\ |t_{j}-t_{k}|\leq 2^{-l} \\ \mbox{for all
          $j,k>h$.} \end{array}\end{equation}  The proof that the
          set of equivalence classes of Cauchy sequences are real
          numbers requires proving that it is a complete ordered field.

          A similar situation holds for rational numbers (and
          integers and natural numbers). Rational numbers are
          elements of any set that satisfies the axioms for an
          ordered field.  However the field is not complete.

          The representation of rational numbers used here will be
          based on  finite strings of digits or kits in
          some base $k$ along with a sign and $"k-al"$ point.
          Use of the string representation is based on the fact that
          physical representations of rational numbers (and integers and
          natural numbers) are given as digits or states of strings
          of kits or qukits (with a sign and $"k-al"$ point for rational
          numbers) in some base $k\geq 2$.  Such numbers are also the base
          of all computations, mainly because of the efficiency in
          representing large numbers and in carrying out arithmetic
          operations. The usefulness of this representation is based
          on the fact that for any base $k\geq 2,$ these string numbers are
          dense in the set of all rational numbers.

          Here, to keep things simple, representations will be
          limited to binary ones with $k=2.$ The representations will
          be further restricted here to be represented as states of
          finite strings of qubits.  This is based on the fact that
          quantum mechanics is the
          basic mechanics applicable to all physical systems.  The
          Cauchy condition will be applied to sequences of these
          states to give quantum theory representations of real
          numbers.

          It should be noted that there are also quantum theory
          representations of real numbers that are not based on
          Cauchy sequences of states of qukit strings. Besides
          those described in \cite{Litvinov,Corbett,Tokuo} there
          are representations as Hermitian operators in Boolean
          valued models of ZF set theory \cite{Takeuti,Davis,Gordon}
          and in a category theory context \cite{Krol}.  These
          representations will not be considered further here because
          of the limitation here of representations to those based on finite
          strings of  qubits.

          \subsection{Rational Number States}

          Here a compact representation of rational string states is used
          that combines the location of the $"binal"$ point and the sign.
          For example,  the state $|1001-0111\rangle$ is a
          state of eight $0-1$ qubits and one $\pm$ qubit representing a
          rational string number $-9.4375$ in the ordinary decimal
          form.

          Qubit strings and their states can be described by locating
          qubits on an integer lattice\footnote{Note that
          the only relevant properties of the integer locations is their
          ordering, a discrete linear ordering.  Nothing is assumed
          about the spacing between adjacent locations on the lattice.}
          Rational string states correspond
          to states of qubits occupying an integer interval $[m+l,m+u].$
          Here $l\leq 0\leq u$, $m$ is the location of the $\pm$ qubit,
          and the $0-1$ qubits occupy all positions in the interval
          $[m+l,m+u].$  Note that two qubits, a sign one and a
          $0-1$ one, occupy position $m.$ For fermionic qubits this
          can be accounted for by including extra variables to
          distinguish the qubit types.

          One way to describe rational string states is by strings
          of qubit annihilation creation (AC) operators $a_{\a,j},\ad_{\a,j}$
          acting on a qubit vacuum state $|0\rangle.$ Also present
          is another qubit type represented by the AC operators
          $c_{\g,m}\cd_{g,m}$. Here $\a =0,1;\g=+,-$, and $j,m$ are
          integers.

          For this work it is immaterial whether the AC operators satisfy
          commutation relations or anticommutation relations:
          \begin{equation}\label{acomm} \begin{array}{c}[\aj,\adjp]=
          \delta_{j,j^{\p}}\delta_{\a,\ap} \\
         \mbox{$[\adj,\adjp]=[\aj,\ajp]=0$}\end{array}\end{equation}or
         \begin{equation}\label{aacomm} \begin{array}{c}\{\aj,\adjp\}=
         \delta_{j,j^{\p}} \delta_{\a,\ap} \\
         \{\adj,\adjp\}=\{\aj,\ajp\}=0.\end{array}\end{equation}
         with similar relations for the $c$ operators. The $c$
         operators commute with the $a$ operators.

         Rational number states are represented by strings of $a$
         creation operators  and one $c$ creation operator acting
         on the qubit vacuum state $|0\rangle$ as \begin{equation}
         \label{rastrst}|\g,m,s,l,u\rangle=\cd_{\g,m}
         \ad_{s(u),u}\cdots\ad_{s(l),
         l}|0\rangle.\end{equation} Here $l\leq m\leq
         u$ and $l<u$ with $l,m,u$ integers, and $s:[l,u]\rightarrow
         \{0,1\}$ is a $\{0,1\}$ valued function on the integer
         interval $[l,u]$. Alternatively $s$ can
         be considered as a restriction to $[l,u]$ of a function
         defined on all the integers.

         An operator $\tilde{N}$ can be defined whose eigenvalues
         correspond to the values of the rational numbers one associates
         with the string states. $\tilde{N}$ is the product of two
         commuting operators, a sign scale operator $\tilde{N}_{ss}$,
         and a value operator $\tilde{N}_{v}.$ One has\begin{equation}
         \label{defN} \begin{array}{c}\tilde{N}=\tilde{N}_{ss}
         \tilde{N}_{v} \\ \mbox{where } \tilde{N}_{ss} =\sum_{\g,m}\g
         2^{-m}\cd_{\g,m}c_{\g,m} \\ \tilde{N}_{v}=\sum_{i,j}i2^{j}
         \ad_{i,j}\a_{i,j}.\end{array}\end{equation} The operator is
         given for reference only as it is not used to define
         arithmetic properties of the rational string states.

         There is a large amount of arithmetical redundancy
         in the states. For instance the arithmetic properties of a
         rational string state are invariant under a translation
         along the integer axis.  This is a consequence of the fact
         that these properties of the state are determined
         by the distribution of $1s$ relative to the position $m$
         of the sign and not on the value of $m$. The other
         redundancy arises from the fact that states that differ
         in the number of leading or trailing $0s$ are all
         arithmetically equal.

         These redundancies can be used to define equivalence classes
         of states or select one member as a representative of each
         class.  Here the latter choice will be made in that rational number
         states will be restricted to those with  $m=0$ for the sign
         location and those with $s$ restricted so that $s(l)=1$ if
         $l<0$ and $s(u)=1$ if $u>0.$ This last condition removes
         leading and trailing $0s.$ The state
         $\ad_{0,0}\cd_{+,0}|0\rangle$ is the number $0$. For ease
         in notation from now on the variables $m,l,u$ will be
         dropped from states. Thus states $|0,\g,s,l,u\rangle$ will
         be represented as $|\g,s\rangle$ with the values of $l,u$
         included in the definition of $s$.

         There are two basic arithmetic properties, equality, $=_{A}$
         and ordering $\leq_{A}$ Arithmetic equality is defined by
         \begin{equation}\label{equalA}\begin{array}{c}
         |\g,s\rangle =_{A}|\g^{\p},s^{\p}\rangle,
         \\ \mbox{ if } l^{\p}=l,\; u^{\p}=u,\;
         \g^{\p}=\g \ \mbox{ and } 1_{s^{\p}}=1_{s}.\end{array}
         \end{equation}  Here $1_{s}=\{j:s(j)=1\}$  the set
         of integers $j$ for which $s(j)=1$ and similarly for $1_{s^{\p}}.$
         That is, two states are arithmetically equal if one has the same
         distribution of $1s$ relative to the location
         of the sign as the other.

         Arithmetic ordering on positive rational string states is
         defined by \begin{equation}\label{deforderA}
        |+,s\rangle \leq_{A}|+,s^{\p}\rangle,
         \mbox{ if } 1_{s}\leq 1_{s^{\p}} \end{equation}
         where\begin{equation}\label{deforderAone}
         \begin{array}{c}1_{s}< 1_{s^{\p}}\mbox{ if there is a $j$
         where $j$ is in $1_{s^{\p}}$ and not in
         $1_{s}$} \\ \mbox{and for all $m>j, m\epsilon 1_{s}$
         iff $m\epsilon 1_{s^{\p}}$}.\end{array}
         \end{equation}  The extension to zero and negative
          states is given by\begin{equation}\label{0negordr}
         \begin{array}{c}|+,\underline{0}\rangle \leq_{A}|+,s\rangle
         \mbox{ for all  $s$}\\ |+,s\rangle \leq_{A}|+,s^{\p}\rangle
         \rightarrow |-,s^{\p} \rangle\leq_{A}|-,s\rangle.
         \end{array}\end{equation}

         The definitions of $=_{A},\leq_{A}$ can be extended to
         linear superpositions of rational string states in a
         straightforward manner to give probabilities that two states are
         arithmetically equal or that one state is less than another
         state.

         Operators for the basic arithmetic operations of addition,
         subtraction, multiplication, and division to
         any accuracy $|+,-\ell\rangle$ are represented by
          $\tilde{+}_{A},\tilde{-}_{A},\tilde{\times}_{A},
          \tilde{\div}_{A,\ell}.$  The state
          \begin{equation}\label{accur}|+,-\ell\rangle
         =\cd_{+,0}\ad_{\underline{0}_{[0,-\ell+1]}}\ad_{1,-\ell}|0\rangle.
         \end{equation} where $\ad_{\underline{0}_{[0,-\ell+1]}}
         =\ad_{0,0}\ad_{0,-1}\cdots\ad_{0,-\ell+1},$ is an eigenstate
         of $\tilde{N}$ with eigenvalue $2^{-\ell}.$

         As an example of the explicit action of the arithmetic
         operators, the unitary addition operator $\tilde{+}$
         satisfies \begin{equation}\label{addn}
         \tilde{+}_{A}|\g s\rangle|\gp\sp\rangle=|\g
         s\rangle|\g^{\p\p}s^{\p\p}\rangle\end{equation} where
         $|\g^{\p\p}s^{\p\p}\rangle$ is the resulting addend state.
         It is often useful to write the addend state as
         \begin{equation}\label{addnplA} |\g^{\p\p}s^{\p\p}\rangle=
         |\gp\sp+_{A}\g s\rangle =_{A}|\gp,\sp\rangle +_{A}|\g,s\rangle.
         \end{equation} More details on the arithmetic operations
         are given elsewhere \cite{BenRRCNQT}. Note that these
         operations are quite different from the usual quantum theory
         superposition, product, etc. operations. This is the reason
         for the presence of the subscript A.

         \subsection{The Cauchy Condition}

         The arithmetic operators can be used to define rational
         number properties of rational string states and their
         superpositions. They can also be used to define the
         Cauchy condition for a sequence of rational string states.
         Let $\{|\g_{n},s_{n}\rangle :n=1,2,\cdots\}$ be any sequence
         of rational string states. Here  for each $n$ $\g_{n}\epsilon
         \{+,-\}$ and $s_{n}$ is a $0-1$ valued function from a finite
         integer interval that includes $0.$

         The sequence $\{|\g_{n}s_{n}\rangle\}$
         satisfies the Cauchy condition if \begin{equation}\label{cauchy}
          \begin{array}{c}\mbox{ For each $\ell$ there is an $h$
          where for all $j,k>h$} \\
          |(|\g_{j}s_{j}-_{A}\g_{k}s_{k}|_{A})\rangle
          <_{A}|+,-\ell\rangle.\end{array} \end{equation} In this
          definition $|(|\g_{j}s_{j}-_{A}\g_{k}
          s_{k}|_{A})\rangle$ is the state that is
          the arithmetic absolute value of the arithmetic difference
          between the states $|\g_{j},s_{j}\rangle$ and
          $|\g_{k},s_{k}\rangle.$ The Cauchy condition says that
          this state is arithmetically less than or equal to the
          state $|+,-\ell\rangle$ for all $j,k$ greater than some $h$.

          It must be emphasized that this Cauchy condition statement
          is a direct translation of Eq. \ref{cauchynos} to apply to
          rational string states.  It has nothing to do with the
          usual convergence of sequences of states in a Hilbert or
          Fock space. It is easy to see that state sequences which
          converge arithmetically do not converge quantum
          mechanically.

          It was also seen in \cite{BenRRCNQT} that the Cauchy condition
          can be extended to sequences of linear superpositions of
          rational states. Let $\psi_{n}
          =\sum_{\g,s}|\g, s\rangle\langle\g s|\psi_{n}\rangle.$
          Here $\sum_{s}=\sum_{l\leq 0}\sum_{u\geq 0}\sum_{s_{[l,u]}}$
          is a sum over all integer intervals $[l,u]$ and
          over all $0-1$ valued functions from $[l,u].$ From this
          one can define the probability that the arithmetic absolute value
          of the arithmetic difference between $\psi_{j}$ and
          $\psi_{k}$ is arithmetically less than or equal to
          $|+,-\ell\rangle$ by\begin{equation}\label{Pjml}
          \begin{array}{l}P_{j,m,\ell}= \sum_{\g,s}
          \sum_{\gp,\sp} |\langle\g, s|\psi_{j}\rangle
          \langle\gp, \sp|\psi_{m}\rangle|^{2} \\
          \hspace{2cm} |(|\g,s-_{A}\gp ,\sp|_{A})\rangle \leq_{A}
          |+,-\ell\rangle. \end{array}\end{equation}
          The sequence $\{\psi_{n}\}$ satisfies the Cauchy
          condition if $P_{\{\psi_{n}\}} =1$
          where\begin{equation}\label{limPjkl}
          P_{\{\psi_{n}\}} =\liminf_{\ell\rightarrow\infty}
          \limsup_{h\rightarrow\infty}\liminf_{j,k>h}P_{j,m,\ell}.
          \end{equation} Here $P_{\{\psi_{n}\}}$ is the probability
          that the sequence $\{\psi_{n}\}$ satisfies the Cauchy condition.

          Cauchy sequences can be collected into equivalence classes
          by defining $\{|\g_{n},s_{n}\rangle\} \equiv
          \{|\g^{\p}_{n}
          s^{\p}_{n}\rangle\}$ if the Cauchy condition holds with
          $\g^{\p}_{k}$ replacing $\g_{k}$ and
          $s^{\p}_{k}$ replacing $s_{k}$ in
          Eq. \ref{cauchy}. To this end let $[\{|\g_{n},
          s_{n}\rangle\}]$ denote the equivalence class containing
          the Cauchy sequence $\{|\g_{n}, s_{n}\rangle\}.$ Similarly
          $\{\psi_{n}\}\equiv\{\psi^{\p}_{m}\}$ if
          $P_{\{\psi_{n}\}\equiv\{\psi^{\p}_{m}\}}=1$ where
          $P_{\{\psi_{n}\}\equiv\{\psi^{\p}_{m}\}}$ is given by
          Eqs. \ref{Pjml} and \ref{limPjkl} with $\psi^{\p}_{k}$
          replacing $\psi_{k}$ in Eq. \ref{Pjml}.

          The definitions of $=_{A},\leq_{A},\tilde{+}_{A},\tilde{-}_{A},
          \tilde{\times}_{A},\tilde{\div}_{A,\ell}$ can be lifted to
          definitions of $=_{R},\leq_{R},\tilde{+}_{R},\tilde{-}_{R},
          \tilde{\times}_{R},\tilde{\div}_{R}$ on the set $[\{|\g_{n},
          s_{n}\rangle\}]$ of all equivalence classes. It can be
          shown \cite{Ben RRCRNQT} that $[\{|\g_{n},
          s_{n}\rangle\}]$ with these operations and relations
          is a representation or model of the real number axioms. In
          this sense it is entirely equivalent to $R$ which is the
          real number component of the complex number base for the
          Hilbert space containing the rational string number states that
          were used to define $[\{|\g_{n},s_{n}\rangle\}].$

          Another representation of real numbers can be obtained by
          replacing the sequences $\{|\g_{n},s_{n}\rangle\}$ by
          operators.  This can be achieved by replacing each index
          $n$ by the  rational string state that corresponds to
          the natural number $n$.  These are defined by
          $|\g,s\rangle$ where $\g=+$ and $l=0$ where $l$ is the
          lower interval bound, for the domain of $s$ as a $0-1$
          function over the integer interval $[l,u].$

          In this case each sequence $|\g_{n},
          s_{n}\rangle$ corresponds to an operator $\tilde{O}$
          defined on the domain of natural number states.  One has
            \begin{equation}\label{defO} \tilde{O}|+,s\rangle =
            |\g_{n},s_{n}\rangle \end{equation} where $n$
            is the $\tilde{N}$ (defined in Eq. \ref{defN})
            eigenvalue of the state $|+,s\rangle.$ $\tilde{O}$ is
            defined to be Cauchy if the righthand sequence in Eq.
            \ref{defO} is Cauchy. One can also give a Cauchy
            condition for $\tilde{O}$ by replacing the natural
            numbers in the definition quantifiers by natural number
            states.

            One can repeat the description of equivalence classes of
            Cauchy sequences of states for the operators to obtain
            another representation of real numbers as equivalence
            classes of Cauchy operators.  The two definitions are
            closely related as Eq. \ref{defO} shows, and  should
            be equivalent as representations of real number axioms.
            This should follow from the use of Eq. \ref{defO} to
            replace the left hand expression for the right hand
            expression in all steps of the proofs that Cauchy
            sequences of rational string states satisfy the real
            number axioms.

            \section{Gauge Transformations}\label{GT}
            The representation of rational string states as states
            of strings of qubits as in Eq. \ref{rastrst} implies a
            choice of quantization axes for each qubit. Usually one
            assumes the same axis for each qubit where the axis is
            fixed by some external field. However this is not
            necessary, and in some cases, such as quantum cryptography
            \cite{QCrypt}, rotation of the axis plays an important
            role.

            In general there is no reason why the axes cannot be
            arbitrarily chosen for each qubit. This freedom of
            arbitrary directions for the axes of each qubit
            corresponds to the set of possible local and global
            gauge transformations of the qubit states. Each gauge
            transformation corresponds to a particular choice in
            that it defines the axis direction of a qubit relative
            to that of its neighbors.

            Here a gauge transformation $U$ can be defined as an
            $SU(2)$ valued function on the integers, $U:\{\cdots
            -1,0,1,\cdots\}\rightarrow SU(2).$ $U$ is global if
            $U_{j}$ is independent of $j$, local if it depends on
            $j$. The effect of $U$ on a rational string state
            $|\g,s\rangle$ is given by \begin{equation}\label{Urast}
            U|\g,s\rangle =U_{0}\cd_{\g,0}U_{u}\ad_{s(u),u}\cdots
            U_{l}\ad_{s(l),l}|0\rangle =(\cd_{U_{0}})_{\g,0}
            (\ad_{U_{u}})_{s(u),u}\cdots(\ad_{U_{l}})_{s(l),l}|0\rangle
            \end{equation}
            where\begin{equation}\label{adU}\begin{array}{c}
            (\ad_{U_{j}})_{i,j}=U_{j}\ad_{i,j}=\sum_{k}(U_{j})_{i,k}
            \ad_{k,j}\\ a_{U_{j}})_{h,j}=a_{h,j}U_{j}^{\dag}=
            \sum_{i}(U_{j})^{*}_{i,h}a_{i,j}\end{array}\end{equation}
            These results are based on the representation of $U_{j}$
            as \begin{equation}\label{Uexpand} U_{j}=\sum_{i,h}
            (U_{j})_{i,h}\ad_{i,j}a_{h,j}\end{equation}

            Arithmetic relations and operators transform in the
            expected way. For the relations one defines
         $=_{A,U}$ and $\leq_{A,U}$ by\begin{equation}\label{=AU}
         \begin{array}{c}=_{A,U}:= (U=_{A}U^{\dag}) \\
         \leq_{A,U}:= U\leq_{A} U^{\dag}.
         \end{array}\end{equation} These relations express the fact
         that $U|\g s\rangle =_{AU}U|\gp\sp\rangle$ if and only if
         $|\g,s\rangle =_{A}|\gp,\sp\rangle$ and $U|\g,s\rangle
         \leq_{AU}U|\gp,\sp\rangle$ if and only if
         $|\g,s\rangle\leq_{A}|\gp,\sp\rangle.$

         For  the operation $\tilde{+}_{A}$ one defines $\tilde{+}_{A,U}$
         by \begin{equation}\label{addAU}\tilde{+}_{A,U}:=(U\times U)
         \tilde{+}_{A}(U^{\dag}\times U^{\dag}).\end{equation} Then
         \begin{equation}\label{addAUA}\begin{array}{l}
         \tilde{+}_{A,U}(U|\g,s\rangle\times U|\gp,\sp\rangle)\\ \hspace{1cm}
         =(U\times U) \tilde{+}_{A}(|\g,s\rangle\times |\gp,\sp\rangle)
         \end{array}\end{equation} as expected. This is consistent with the
         definition of $\tilde{+}_{A}$ in Eq. \ref{addn} as a binary relation.
         Similar relations hold for $\tilde{\times}_{A},\tilde{-}_{A},
         \tilde{\div}_{A,l}.$

         It follows from these properties that the Cauchy condition
         is preserved under gauge transformations.  If a sequence of
         states $\{\psi_{n}\}$ is Cauchy then the sequence
         $\{U\psi_{n}\}$  is U-Cauchy  which means that it is Cauchy
         relative to the transformed arithmetic relations and
         operations. For example, if a sequence $\{|\g_{n},
         s_{n}\rangle\}$ satisfies the Cauchy condition,
         then  the transformed sequence $\{U|\g_{n},
         s_{n}\rangle\}$ satisfies the U-Cauchy condition:
         \begin{equation}\label{cauchyU}
          \begin{array}{c}\mbox{ For each $\ell$ there is an $h$
          where for all $j,k>h$} \\|(U|\g_{j}s_{j}-_{AU}
          U\g_{k}s_{k}|_{AU})\rangle
          <_{AU}U|+,-\ell\rangle.\end{array} \end{equation}

          These definitions and considerations extend to the Cauchy
          operators.  If $\tilde{O}$ is Cauchy the the above shows
          that \begin{equation}\label{defOU}\tilde{O}_{U}=
          U\tilde{O}U^{\dag} \end{equation} is U-Cauchy. However
          $\tilde{O}_{U}$ is not a Cauchy operator in the original
         frame and $\tilde{O}$ is not Cauchy in the transformed frame.

         To see that $\tilde{O}_{U}$ is not Cauchy in the original frame
         it is instructive to consider a simple example.
         First one works with the Cauchy
         property for sequences of states. Let
         $f:(-\infty,n]\rightarrow \{0,1\}$ be a $0-1$ function from
         the set of all integers $\leq n$ where $f(n)=1.$ Define a
         sequence of states\begin{equation}\label{fseq}
         |f\rangle_{m} =\cd_{+,0}\ad_{f(n),n}\ad_{f(n-1),n-1}\cdots
         \ad_{f(-m),-m}|0\rangle\end{equation}for $m=1,2,\cdots.$
         The sequence is Cauchy as $||f(j)\rangle
         -_{A}|f(k)\rangle|_{A}\leq_{A}|+,-\ell\rangle$ for all
         $j,k>\ell.$ However for any  gauge transformation $U$
         the sequence \begin{equation}\label{fseqU}
         U|f\rangle_{m} =\cd_{+,0}(\ad_{U})_{f(n),n}\cdots
         (\ad_{U})_{f(-m),-m}|0\rangle\end{equation} is not Cauchy
         as  expansion of the $\ad_{U}$ in terms of the $\ad$ by Eq.
         \ref{adU} gives $U|f\rangle_{m}$ as a sum of terms whose
         arithmetic divergence is independent of $m$.

         To show that $\tilde{O}_{U}$ is Cauchy in the
         transformed frame if and only if $\tilde{O}$ is Cauchy in
         the original frame one can start with the expression for
         the Cauchy condition in the transformed frame:
         \begin{equation}\label{cauchtOU}
         \begin{array}{c}|\tilde{O}_{U}U|s_{j}\rangle
         -_{AU}\tilde{O}_{U}U|s_{k}\rangle|_{AU}\leq_{AU}U|+,-\ell\rangle \\
         \mbox{for all $U|s_{j}\rangle,$ $U|s_{k}\rangle \geq_{AU}$ some
         $U|s_{h}\rangle$}\end{array}\end{equation} From Eq. \ref{defOU}
         one gets $$\tilde{O}_{U}U|s_{j}\rangle
         -_{AU}\tilde{O}_{U}U|s_{k}\rangle=U\tilde{O}|s_{j}\rangle
         -_{AU}U\tilde{O}|s_{k}\rangle.$$ From Eqs. \ref{addn} and
         \ref{addnplA} applied to $-_{AU}$ and Eqs. \ref{addAU} and
         \ref{addAUA} one obtains $$U\tilde{O}|s_{j}\rangle
         -_{AU}U\tilde{O}|s_{k}\rangle =U(\tilde{O}|s_{j}\rangle
         -_{A}\tilde{O}|s_{k}\rangle).$$  Use of\begin{equation}\label{absUA}
         |-|_{AU}=U|-|_{A}U^{\dag} \end{equation}for the absolute
         value operator gives $$|U(\tilde{O}|s_{j}\rangle
         -_{A}\tilde{O}|s_{k}\rangle)|_{AU}=U(|\tilde{O}|s_{j}\rangle
         -_{A}\tilde{O}|s_{k}\rangle|_{A}).$$ Finally from Eq.
         \ref{=AU} one obtains $$\begin{array}{l}U(|\tilde{O}|s_{j}\rangle
         -_{A}\tilde{O}|s_{k}\rangle|_{A})\leq_{AU}U|+,-\ell\rangle
         \\ \hspace{1cm} \leftrightarrow |\tilde{O}|j\rangle
         -_{A}\tilde{O}|s_{k}\rangle|_{A}\leq_{A}|+,-\ell\rangle
         \end{array}$$ which is the desired result. Thus one sees
         that the Cauchy property is preserved in unitary transformations
         from one reference frame to another.

          As was done with the Cauchy sequences and operators, the
          U-Cauchy sequences or their equivalents, U-Cauchy operators,
          can be collected into a set ${\mathcal R}_{U}$ of
          equivalence classes that represent the real numbers.
          This involves lifting up the basic arithmetic relations
          $=_{AU},\leq_{AU}$ and operations $\tilde{+}_{AU},
          \tilde{\times}_{AU},\tilde{-}_{AU}, \tilde{\div}_{AU,l}$
          to real number relations $=_{RU},\leq_{RU}$ and operations
          $\tilde{+}_{RU}, \tilde{\times}_{RU},\tilde{-}_{RU},
          \tilde{\div}_{RU},$ and showing that ${\mathcal R}_{U}$
          is a complete, ordered, field.

         It is also the case that for almost all gauge $U$ the real
         numbers in $\mathcal{R}_{U}$ are orthogonal to those in
         $\mathcal R$ in the following sense.  One can see that each
         equivalence class in $\mathcal R$ contains a state sequence
         $|\g_{n},s_{[u,-n]}\rangle$ where $s$ is a $0-1$ valued function
         on the interval of all integers $\leq u.$ Let $U$ be a gauge
         transformation with associated state sequence
         $|U_{0}\g_{n},Us_{[u,-n]}\rangle.$ Both sequences satisfy
         their respective Cauchy conditions.  However the overlap
         $\langle\g_{n},s_{[u,-n]}|U_{0}\g_{n},Us_{[u,-n]}\rangle
         \rightarrow 0$ as $n\rightarrow\infty.$  This expresses the
         sense in which $\mathcal R$ and $\mathcal{R}_{U}$ are
         orthogonal.

         \section{Fields of Quantum Frames}

            As has been seen, one can
          define many quantum theory representations
          of real numbers as Cauchy sequences of states of qubit
          strings or as Cauchy operators on the qubit string
          states. The large number of representations stems from the
          gauge (global and local) freedom in the choice of a
          quantization axis for the qubit strings. Complex number
          representations are included either as ordered pairs of
          the real number representations or as extensions of the
          description of Cauchy sequences or Cauchy operators to
          complex rational string states \cite{BenRRCNQT}.

          It was also seen that for each gauge transformation $U$
          the real and complex number representations ${\mathcal
          R}_{U},\; \mathcal{C}_{U}$ are the base of a frame $F_{U}$.
          The frame $F_{U}$ also contains representations of
          physical theories as mathematical structures based on
          ${\mathcal R}_{U},\; \mathcal{C}_{U}.$

          The work in the last two sections shows that the description
          of rational string states as states of finite strings of
          qubits given is a description in a Fock space. (A Fock space
          is used because of the need to describe states
          of variable numbers of qubits and their linear
          superpositions in one space.)  The arithmetic operations
          $+_{A},-_{A},\times_{A},\div_{A,\ell}$ on states of
          these strings are represented by Fock space operators.
          The properties of these operators acting on the qubit
          string states are used to show that the states represent
          binary rational numbers. Finally equivalence classes of
          sequences of these states or of operators that satisfy the
          Cauchy condition are proved to be real numbers
          \cite{BenRRCNQT}.

          The essential point here is that the Fock space, $\mathcal F$,
          and any additional mathematics used to obtain these results
          are based on a set $R,C$ of real and complex numbers.  For
          example, superposition coefficients of basis states are
          in $C$, the inner product is a map from pairs of states to
          $C$, operator spectra are elements of $C$, etc. The space
          time manifold used to describe the dynamics of any
          physical representations of the qubit strings is given by
          $R^{4}$.

          It follows that one can assign a reference frame $F$ to
          $R$ and $C.$ Here $F$ contains all physical and mathematical
          theories that are represented as mathematical structures
          based on $R$ and $C$.  However, unlike the case for the frames
          $F_{U},$ the only properties that $R$ and $C$ have are those
          based on the relevant axioms (complete ordered field for $R$).
          Other than that, nothing is known about how they are
          represented.

          This can be expressed by saying that $R$ and $C$ are
          external, absolute, and given.  This seems to be the usual
          position taken by physics in using theories based on $R$
          and $C$. Physical theories are silent on what properties $R$
          and $C$ may have other than those based on the relevant
          axioms.  However, as has been seen, one can use these
          theories to describe many representations $R_{U}$ and $C_{U}$
          and associated frames $F_{U}$ based on $SU(2)$ gauge
          transformations of the qubit strings.  As noted, for each
          $U,$ $F_{U}$ contains representations of all physical
          theories as mathematical structures over $R_{U},C_{U}$.
          For these frames one can see that  $R_{U}$ and
          $C_{U}$ have additional properties besides those given by
          the relevant axioms. They are also equivalence classes of
          Cauchy sequences $\{U|\g_{n},s_{n}\rangle \}$ or Cauchy
          operators $\tilde{O}_{U}.$

\begin{figure}[h]\begin{center}
           \resizebox{120pt}{120pt}{\includegraphics[230pt,200pt]
           [490pt,460pt]{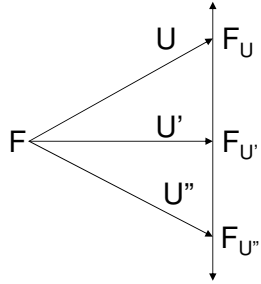}}\end{center}
           \caption{Relation between a Base Frame $F$ and Gauge
           Transformation Frames. A frame $F_{U}$ is associated
           with each gauge transformation $U$ of the rational string
           states in $F$. The three frame connections shown are
           illustrative of the infinitely many connections,
           shown by the two headed vertical arrow. Each $F_{U}$
           is based on real and complex numbers $\mathcal{R}_{U},
           \mathcal{C}_{U}$ and a space time manifold $\mathcal{R}^{4}_{U}$.}
           \label{RCST1} \end{figure}

          Fig. \ref{RCST1} is a schematic illustration of the
          relation between frame $F$ and  the frames $F_{U}$. Only
          three of the infinitely many $F_{U}$ frames are shown.
          The arrows indicate the derivation direction in that $R,C$
          based theory in $F$ is used to describe, for each $U$
          $R_{U}$ and $C_{U}$ that are the base of all theories in
          $F_{U}$. Note that the frame $F_{ID}$ with the
          identity gauge transformation is also included as one of
          the $F_{U}.$  It is \emph{not} the same as $F$ because
          $R_{ID}$ is not the same as $R$.

           The above relations between the frames $F$ and $F_{U}$
           shows that one can repeat the description of real and
           complex numbers as Cauchy sequences of (or Cauchy
           operators on) rational string states in each frame
           $F_{U}$.  In this case the Fock space representation,
           ${\mathcal F}_{U}$, used to describe the qubit string
           states in $F_{U}$, is different from from $\mathcal F$ in
           that it is based on $R_{U},C_{U}$ instead of on $R,C.$
           However the two space representations are related by an
           isomorphism that is based on the isomorphism between
           $R$ and $R_{U}.$

           It is useful to examine this more.  First consider the
           states of a qubit at site j. For an observer
           in frame $F,$ these states have the general form
           $\alpha|0\rangle+\beta|1\rangle$  where $\alpha$ and
           $\beta$ are complex numbers in $C$. Let $U$  be
           an $SU(2)$ gauge transformation where $U(j)$ is defined by
           $$\begin{array}{c}U(j)|0\rangle =|+\rangle \\ U(j)|1\rangle
           =|-\rangle\end{array}$$ where $|\pm\rangle= (1/\sqrt{2})
           |0\rangle \pm |1\rangle).$  Then the states
           $|+\rangle$ and $|-\rangle$ in frame $F$ are seen by an
           observer in $F_{U}$ as the states $|0\rangle$ and
           $|1\rangle$ respectively as the quantization axis is
           different.

           A similar situation holds for states of qubit strings.
           To an observer in $F$ the state $U|\g,s\rangle$ is
           different from $|\g,s\rangle.$  However an observer in
           frame $F_{U},$ with a different set of quantization
           axes for each qubit, would represent the state
           $U|\g,s\rangle$ as $|\g,s\rangle$ as to him it is the
           same state relative to his axis set as it is to the
           observer in $F$ for his axis set.

           The situation is slightly different for linear
           superpositions of basis states. To an observer in $F$,
           the coefficients $\alpha,\beta$ in the state $\alpha U|
           \g,s\rangle +\beta U|\gp,\sp\rangle$  represent abstract
           elements of $C$.  The same observer sees that this state
           in $F_{U}$ is represented by $\alpha_{U}|\g,s\rangle
           +\beta_{U}|\gp,\sp\rangle$ where $\alpha_{U}, \beta_{U}$
           as elements of $C_{U}$, represent the same abstract
           complex numbers as do $\alpha,\beta.$  However an observer
           in $F_{U}$ sees this same state as $\alpha|\g,s\rangle
           +\beta|\gp,\sp\rangle.$ To him the real and complex
           number base of his frame is abstract and is represented by $R,C.$

           In general an observer in any frame sees the real
           and complex number base of his own frame as abstract and given with no
           particular representation evident. However the observer
           in frame $F$ also sees that what to him is the abstract
           number $r,$ is the number $r_{U}$ as and element of
           $R_{U}$ in frame $F_{U}.$

           These considerations also extend to group representations as
           matrices of complex numbers. If the element $g,$ as an
           abstract element of $SU(2),$ is represented in frame $F$
           by the matrix $\left| \begin{array}{cc} a & b\\c & d
           \end{array}\right|$ where $a,b,c,d$ are elements of $C$,
           then, to an observer in $F,$ $g$ is represented in frame
           $F_{U}$ by $\left| \begin{array}{cc} a_{U} & b_{U}\\c_{U} &
           d_{U}\end{array}\right|.$ Here $a_{U},b_{U},c_{U},
           d_{U}$, as elements of $C_{U}$, correspond to the same
           abstract complex numbers as do $a,b,c,d.$  However an
           observer in $F_{U}$ sees this representation as
           $\left| \begin{array}{cc} a & b\\c & d
           \end{array}\right|$ which is the same as the $F$ observer
           sees in his own frame.

\begin{figure}[h]\begin{center}
           \resizebox{130pt}{130pt}{\includegraphics[250pt,160pt]
           [540pt,450pt]{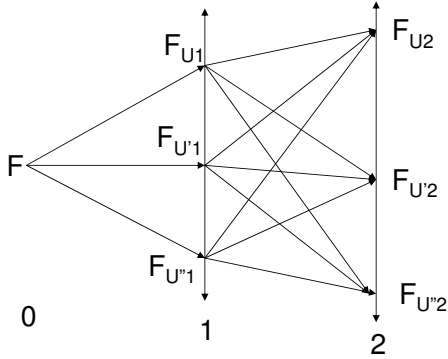}}\end{center}
           \caption{Three Iteration Stages of Frames coming from
           Frames. Only three frames of the infinitely many, one for
           each gauge $U,$ are shown at stages $1$ and $2$. The
           arrows connecting the frames show the iteration direction
           of frames emanating from frames.} \label{RCST2}
           \end{figure}

           Following this line of argument one can now describe
           another generation of frames with each frame $F_{U}$ in
           the role of a parent to progeny frames just as $F$ is a
           parent to the frames $F_{U}$ as in Fig. \ref{RCST1}.
           This is shown schematically in Fig. \ref{RCST2}. Again,
           only three of the infinitely many stage 2 frames
           emanating from each stage 1 frame are shown.
           Here something new appears in that there are many
           different paths to a stage 2 frame. For each path, such as
           $F\rightarrow F_{U_{1}}\rightarrow F_{U_{2}},$ $U_{2}$ is
           the product $U^{\p\p}U_{1}$ of two gauge transformations
           where $U^{\p\p}=U_{2}U^{\dag}_{1}.$  An observer in this
           frame $F_{U_{2}}$ sees the real and complex number frame
           base as abstract, and given.  To him they can be
           represented as $R,C.$ An observer in $F_{U_{1}}$ sees
           the real and complex number base of $F_{U_{2}}$ as
           $R_{U^{\p\p}},C_{U^{\p\p}}.$

           However an observer in $F$ sees the
           real and complex number base of $F_{U_{2}}$ as
           $R_{U^{\p\p}|U_{1}},C_{U^{\p\p}|U_{1}}.$ The subscript
           $U^{\p\p}|U_{1}$ denotes the fact that relative to $F$
           the number base of $F_{U_{2}}$ is constructed in two
           stages. First the Fock space $\mathcal F$ is used to
           construct representations $R_{U_{1}},C_{U_{1}}$ of
           $R$ and $C$ as $U_{1}$ Cauchy
           sequences of states $\{U_{1}|\g_{n},s_{s}\rangle\}.$
           Then in frame $F_{U_{1}}$ the Fock space ${\mathcal
           F}_{U_{1}},$ based on $R_{U_{1}},C_{U_{1}},$ is used to
           construct the number representation base of $F_{U_{2}}$
           as $U^{\p\p}$ Cauchy sequences $\{U^{\p\p}|\g_{n},s_{n}
           \rangle\}$ of qubit string states in ${\mathcal F}_{U_{1}}.$

           One sees, then, that, for each path leading to a specific
           stage 2 frame, there is a different two stage construction
           of the number base of the frame.  This view from the
           parent frame $F$ is the same view that we have as observers
           outside the whole frame structure. That is, our
           external view coincides with that for an observer inside
           the parent frame $F$.

\begin{figure}[h]\begin{center}
           \resizebox{130pt}{130pt}{\includegraphics[230pt,120pt]
           [560pt,490pt]{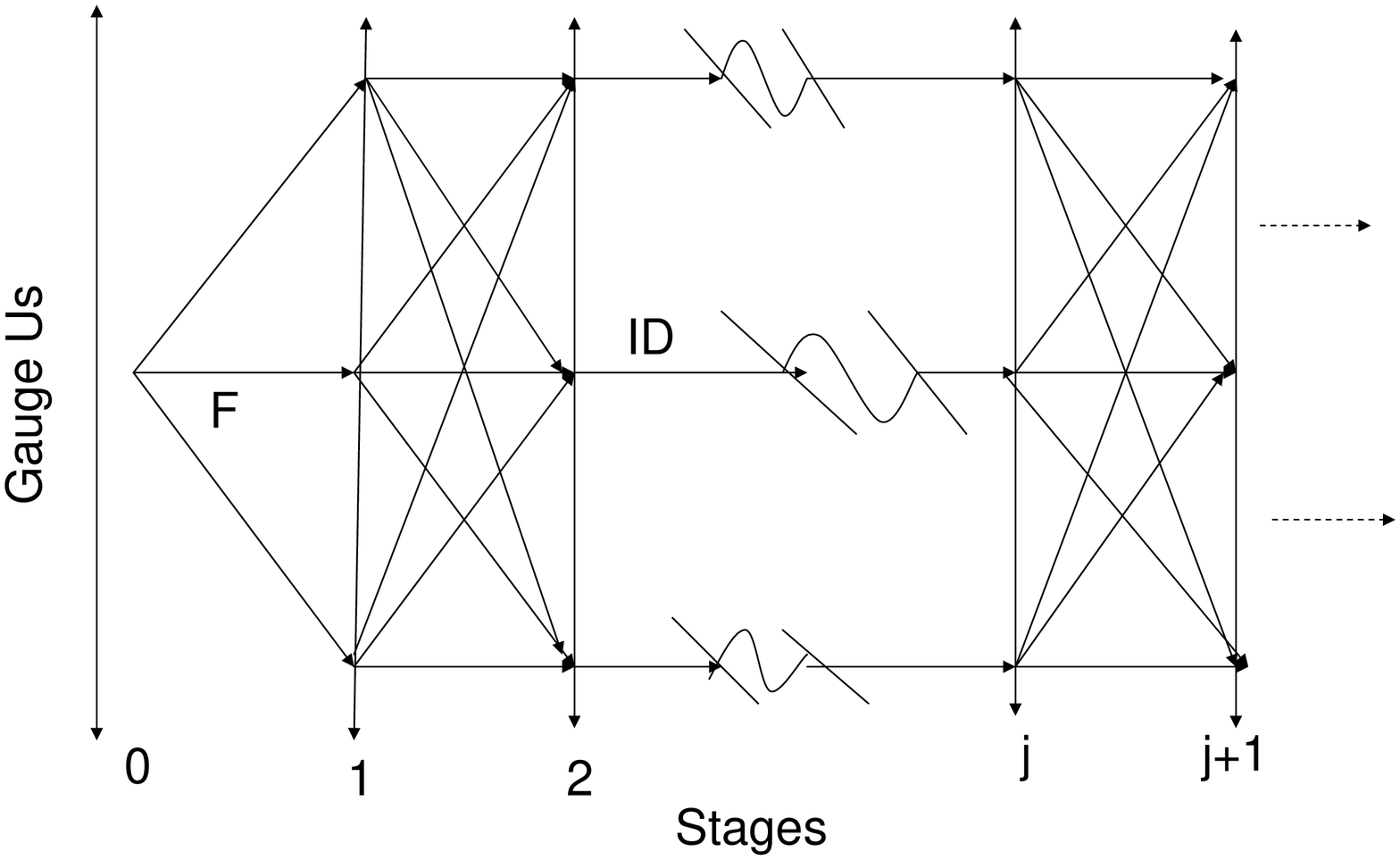}}\end{center}
           \caption{One way Infinite Iteration of Frames coming from
           Frames. Only three of the infinitely many frames, one for
           each gauge $U$ are shown for stages $1,2,\cdots,j,j+1,\cdots.$
           The arrows connecting the frames show the iteration or emanation
           direction. The center arrows labeled ID denote iteration of the
           identity gauge transformation.} \label{RCST3}
           \end{figure}

           The above description of frames emanating from frames for
           2 stages suggests that the process can be iterated.
           There are several possibilities besides a finite number
           of iterations exemplified by Fig.\ref{RCST2} for 2
           iterations.  Fig. \ref{RCST3} shows the field structure
           for a one way infinite number of iterations.
             Here one sees that each frame has an
           infinite number of descendent frame generations and, except for
           frame $F,$ a finite number number of ancestor
           generations. The structure of the frame field seen by an
           observer in $F$ is the same as that viewed from the
           outside.  For both observers the base real and complex numbers
           for $F$ are seen as abstract and given with no structure
           other than that given by the axioms for the real and
           complex numbers.

\begin{figure}[h!]\begin{center}
           \resizebox{130pt}{130pt}{\includegraphics[230pt,120pt]
           [560pt,490pt]{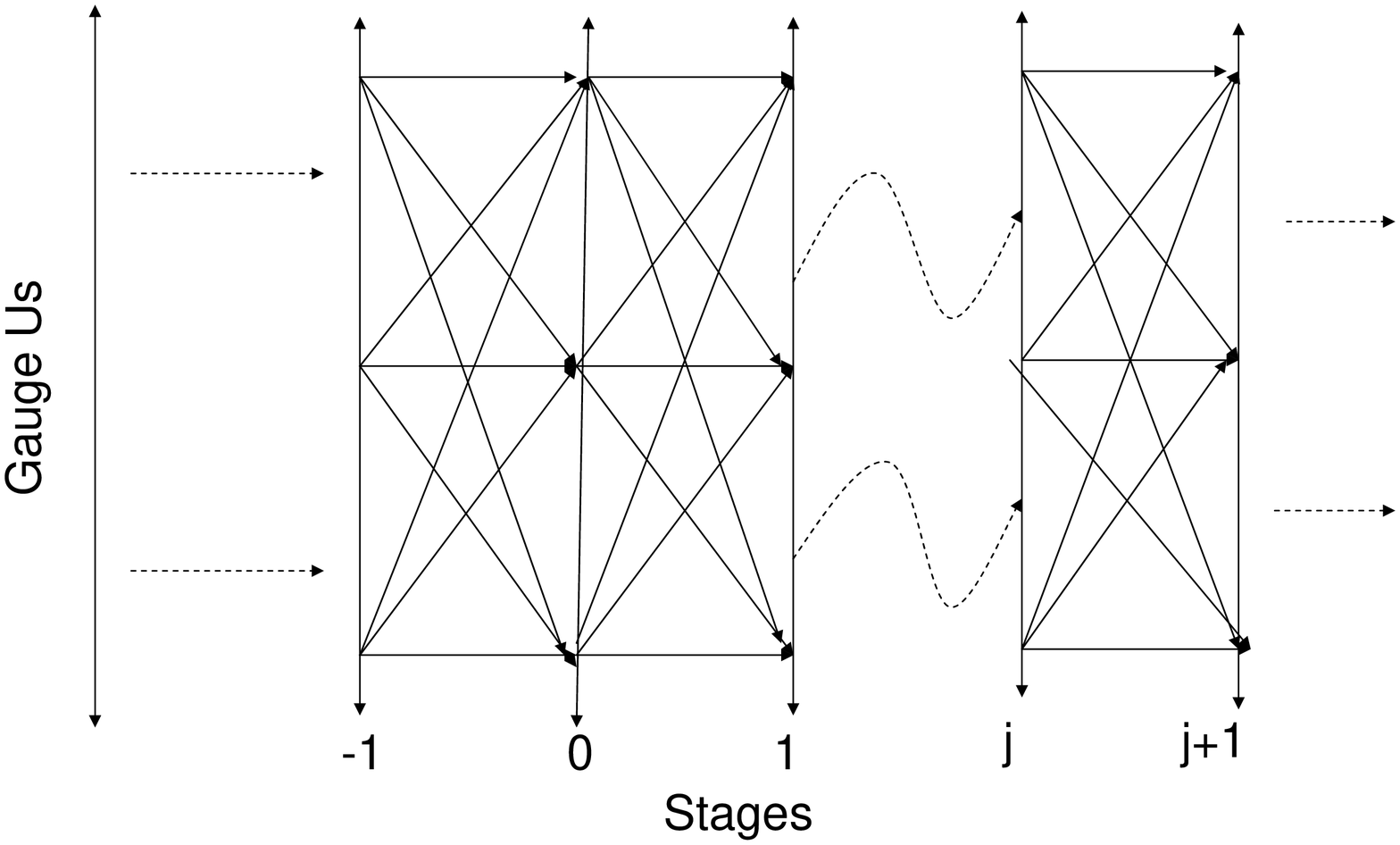}}\end{center}
           \caption{Two way Infinite Iteration of Frames coming from
           Frames. Only three of the infinitely many frames, one for
           each gauge $U$ are shown at each stage $\cdots ,-1,0,1,
           \cdots,j,j+1,\cdots$. The solid arrows connecting the
           frames show the iteration or emanation direction. The wavy
           arrows denote iterations connecting stage $1$ frames to
           those at stage $j$.  The straight dashed arrows denote
           infinite iterations from the left and to the right.of the
           identity gauge transformation.} \label{RCST4} \end{figure}

\begin{figure}[h!]\begin{center}
           \resizebox{130pt}{130pt}{\includegraphics[230pt,120pt]
           [560pt,490pt]{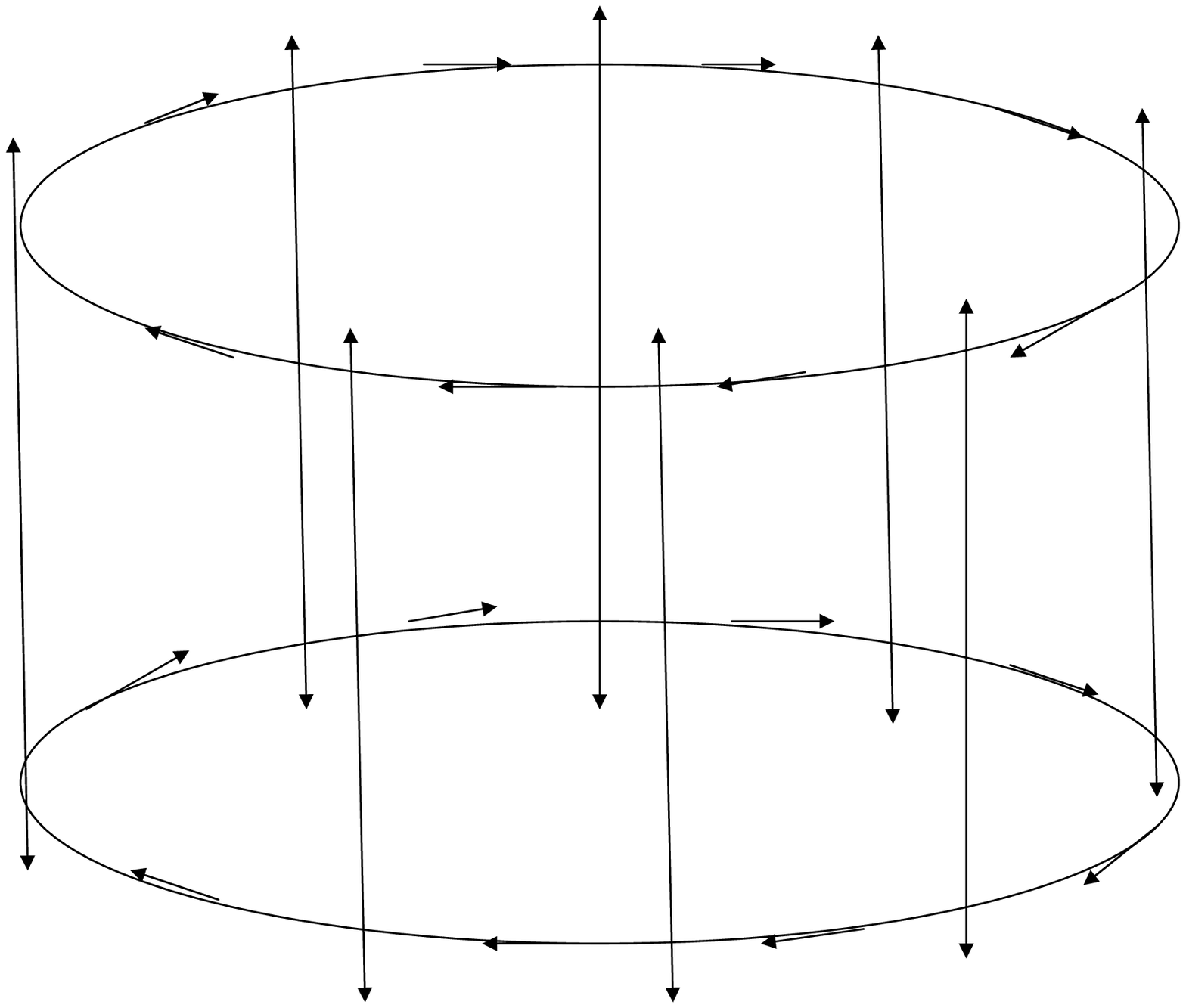}}\end{center} \caption{Schematic
           Representation of Cyclic Iteration of Frames coming from
           Frames. The vertical two headed arrows represent the gauge
           transformations at each stage and the arrows along both
           ellipses show the direction of iteration. To avoid a very
           complex and messy figure no arrows connecting frames to
           frames are shown.} \label{RCST5} \end{figure}

           There are two other possible stage or generation structures
           for the frame fields, two way infinite  and finite cyclic
           structures. These are shown schematically in Figs.
           \ref{RCST4} and \ref{RCST5}.  The direction of
           iteration in the cyclic field is shown by the arrows on
           the circle rather than example arrows connecting frames
           as in the other figures. For both these frame fields each
           frame has infinitely many parent frames and infinitely
           many daughter frames. There is no single ancestor frame
           and no final descendent frames. The cyclic field is
           unique in that each frame is both its own descendent and
           its own ancestor.  The distance between these connections
           is equal to the number of iterations in the cyclic field.

           These two frame fields differ from the others in that the
           structure seen from the outside is different from that for an
           observer in any frame. There is no frame $F$ from which the view is
           the same as from the outside. Viewed from the outside there are no
           abstract, given real and complex number sets for the field as a
           whole.  All of these are internal in that an observer in
           frame $F_{U_{j}}$ at generation $j$ sees the base
           $R_{U_{j}},C_{U_{j}}$ as abstract with no properties
           other than those given by the relevant axioms.

           The same holds for the representations of the space time
           manifold.  Viewed from the outside there is no fixed
           abstract space time representation as a 4-tuple of real numbers
           associated with the field as a whole.  All space time
           representations of this form are internal and associated
           with each frame. This is based on the observation that
           the points of a continuum space time as a 4-tuple of
           representations of the real numbers are different in
           each frame because the real number representations are
           different in each frame.  Also, contrary to the
           situation for the fields in Figs. \ref{RCST1}-\ref{RCST3},
           there is no  representation of the space time points
           that can be considered to be as fixed and external to
           the frame field.

           The lack of a fixed abstract, external space time manifold
           representation for the two-way infinite and cyclic frame
           fields is in some ways similar to the lack of a
           background space time in loop quantum gravity
           \cite{Smolin}.  There are differences in that in
           loop quantum gravity space is discrete on the Planck
           scale and not continuous \cite{Ashtekar}.  It should be
           noted though that the representation of space time as
           continuous is not a necessary consequence of the frame
           fields and their properties.   The important part is the
           real and complex number representation base of each
           frame, not whether space time is represented as discrete
           or continuous.

           It is useful to summarize the views of observers inside
           frames and outside of frames for the different field
           types.  For all the fields except the cyclic ones an
           observer in any frame $F_{U_{j}}$ at stage $j$ sees the real
           number base $R_{U_{j}}$ of his frame as abstract and external
           with no properties other than those given by the axioms for a
           complete ordered field. The observer also cannot see any
           ancestor frames. He/she can see the whole frame
           field structure for all descendent frames at stages
           $k>j.$  Except as noted below, the view of an outside
           observer is different in that he/she can see the whole
           frame field structure. This includes the point that, to
           internal observers in a frame, the real and complex
           number base of the frame is abstract and external.

           For frame fields with a fixed ancestor frame $F$, Figs.
           \ref{RCST1}, \ref{RCST2}, \ref{RCST3}, the view of an outside
           observer is almost the same as that of an observer in frame
           $F$. Both see the real and complex number base of $F$ as
           abstract and external.  Both can also see the field structure
           for all frames in the fields. However the outside observer
           can see that frame $F$ has no ancestors. This is not
           available to an observer in $F$ as he/she cannot see the
           whole frame field.

           The cyclic frame field is different in that for an
           observer in any frame at stage $j,$ frames at other stages
           are both descendent and ancestor frames. This suggests
           that the requirement that a frame observer cannot see
           the field structure for ancestor frames, but can see it for
           descendent frames, may have to be changed, at least for
           this type of frame field.  How and what one learns from
           such changes are a subject for future work.

         \section{Relation between the Frame Field and Physics}
         \label{RFFP} So far, frame fields based on different quantum
         mechanical representations of real and complex numbers
         have been described. Each frame contains a representation
         of physical theories as mathematical structures based
         on the real number representation base of the frame. The
         representations of the physical theories in the different
         frames are different because the real (and complex) number
         representations are different. They are also
         isomorphic because the real (and complex) number
         representations are isomorphic.

         The description of the frame field  given so far is incomplete
         because nothing has been said about the relation of the
         frame field to physics. So far the only use of physics has
         been to limit the description of rational number representations
         to quantum mechanical states of strings of qubits.

         The main problem here is that to date all physical theories
         make no use of this frame field.  This is evidenced by the
         observation that the only properties of real numbers
         relevant to physical theories are those derivable from the
         real number axioms for a complete, ordered field.  So far
         physical theories are completely insensitive to details of
         different representations of the real numbers.

         This problem is also shown by the observation that there is no
         evidence of this frame structure and the multiplicity of
         real number representations in our view of the existence
         of just one physical universe with its space time manifold,
         and  with real and complex numbers that can be treated as
         absolute and given. There is no hint, so far, either in
         physical theories or in properties of  physical systems and
         the physical universe, of the different representations and
         the frame fields.

         This shows that the main problem is to reconcile the
         great number of different representations of the real and
         complex numbers and the $R^{4}$ space time manifold as
         bases for different representations of physical
         theories with the lack of dependence of physical theories
         on these representations and no evidence for them in our
         view of the physical universe.

         One possible way to do this might be to collapse the frame
         field to a smaller field, ideally with just one frame.  As
         a step in this direction one could limit quantum theory
         representations of rational string numbers to those that
         are gauge invariant. This would have the effect of collapsing
         all frames $F_{U_{j}}$ at any stage $j$ into one stage
         $j$ frame\footnote{Note that $U_{j}$ is a gauge
         transformation.  It is not the $jth$ element of one.}.
         The resulting frame field would then be one
         dimensional with one frame at each stage.

         The idea of constructing representations that are gauge
         invariant for some gauge transformations has already been
         used in another context.  This is the use of the decoherent
         free subspace (DFS) approach to quantum error correction.
         This approach \cite{Lidar,LidarII} is based to quantum error
         avoidance in quantum computations.  This method identifies
         quantum errors with gauge transformations $U$. In this case
         the goal is to find subspaces of qubit string states that are
         invariant under at least some gauge $U$ and are preserved by
         the Hamiltonian dynamics for the system.

         One way to achieve this is based on irreducible representations
         of direct products of $SU(2)$ as the irreducible subspaces are
         invariant under the action of some $U$. As an example, one can
         show that \cite{Enk} the subspaces defined by the irreducible
         4 dimensional representation of $SU(2)\times SU(2)$ are invariant
         under the action of any global $U$. The subspaces are the three
           dimensional subspace with isospin $I=1$, spanned by the states
           $|00\rangle,|11\rangle,1/\sqrt{2}(|01\rangle +|10\rangle)$
           and the $I=0$ subspace containing $1/\sqrt{2}(|01\rangle
           -|10\rangle).$  The action of any global $U$ on
           states in the $I=1$ subspace can change one of the
           $I_{z}$ states into linear superpositions of all states in
           the subspace.  But it does not connect the states in the $I=1$
           subspace with that in the $I=0$ subspace.

           It follows that one can replace a string of $2n$  qubits
           with a string of $n$ new qubits where the
           $|0\rangle$ and $|1\rangle$ states of the $jth$
           new qubit correspond to any state in the respective
           $I=1$ and $I=0$ subspaces of the 4 dimensional
           representation of $SU(2)_{2j-1}\times SU(2)_{2j}.$ Any
           state of the $n$ new qubits is invariant under all
           global gauge transformations and all local gauge
           transformations where \begin{equation}\label{Ueq}
           U_{2j-1}=U_{2j}.\end{equation}

           This replacement of  states of strings of $2n$ qubits
           by states of strings of $n$ new qubits gives the result
           that, for all $U$ satisfying Eq. \ref{Ueq}, the $F_{U}$ frames at
           any stage $j$ all become just one frame at stage $j$.
           However this still leaves many gauge $U$ for which the new
           qubit string state representation is not gauge invariant.

          Another method of arriving at a gauge invariant
          description of rational string states is based on the
          description of the kinematics of a quantum system  by
          states  in a Hilbert space $\mathcal H,$ based on the $SU(2)$ group
          manifold.  Details of this, generalized to all compact Lie
          groups, are given in \cite{Mukunda} and \cite{Ashtekar}.
          In essence this extends the well known situation for
          angular momentum representations of states based on the
          manifold of the group $SO(3)$ to all compact Lie groups.

          For the angular momentum case the the action of any
          rotation on the states $|l,m\rangle$ gives linear
          combinations of states with different $m$ values but with
          the same $l$ value.  The Hilbert space spanned by all
          angular momentum eigenstates can be expanded as a
          direct sum \begin{equation}\label{Hl}
          \mathcal H =\bigoplus_{l}\mathcal{H}_{l} \end{equation}
          where $l=0,1,2,\cdots$ labels the irreducible
          representations of $SO(3).$  Qubits can be associated with
          this representation by choosing two different $l$ values,
          say $l_{0}$ and $l_{1}.$ Then any states in the subspaces
          $\mathcal{H}_{l_{0}}$ and $\mathcal{H}_{l_{1}}$ correspond
          to the $|0\rangle$ and $|1\rangle$ qubit states
          respectively. These states are invariant under all
          rotations.  Extension of this construction to all finite qubit
          strings gives a representation of  natural numbers,
          integers and rational numbers that is invariant under all
          $SO(3)$ gauge transformations.

          This development can also be carried out for any compact
          Lie group where the quantum kinematics of a system is
          based on the group manifold \cite{Mukunda,Ashtekar}.
          In the case of $SU(2)$ Eq. \ref{Hl} holds with
          $l=0,1/2,1.3/2,\cdots$. The momentum
          subspaces $\mathcal{H}_{l}$ are invariant under all
          $SU(2)$ transformations. As in the angular momentum
          case one can use this to describe states of qubits as
          \begin{equation}\label{invqub}\begin{array}{c}
          |0\rangle\rightarrow  \mathcal{H}_{l_{0}} \\
          |1\rangle\rightarrow  \mathcal{H}_{l_{0}}\end{array}
          \end{equation} that are $SU(2)$ invariant.

          This construction can be extended to states of finite
          strings of qubits.  Details of the mathematics needed for
          this, applied to graphs on a compact space manifold,
          are given in \cite{Ashtekar}.  In this way one can
          describe representations of rational string numbers that
          are $SU(2)$ gauge invariant for all gauge $U$.

          There is a possible connection between these representations
          of numbers and the Ashtekar approach to loop quantum gravity.
          The Ashtekar approach \cite{Ashtekar} describes $G$ valued
          connections on graphs defined on a $3D$ space manifold where
          $G$ is a compact Lie group. The Hilbert space of states on all
          graphs can be represented as a direct sum of spaces for each graph.
          The space for each graph can be further decomposed into a sum
          of gauge invariant subspaces. This is similar to the spin
          network decomposition first described by Penrose
          \cite{Penrose}.

          The connection to qubit strings is made by noting that
          strings correspond to simple one dimensional graphs.
          States of qubit strings are defined as above by choosing
          two $l$ values for the space of invariant subspaces as in
          Eq. \ref{invqub}.  It is hoped to describe more details
          of this connection in future work.

          Implementation of this approach to reduction of the frame
          field one still leaves a one dimensional line of
          iterated frames. The line is finite, Fig. \ref{RCST2},
          one way infinite, Fig. \ref{RCST3}, two way infinite,
          Fig. \ref{RCST4}, or closed, Fig. \ref{RCST5}. Because the
          two way infinite and cyclic fields have no abstract
          external sets of real and complex numbers and no abstract
          external space time, it seems appropriate to limit
          consideration to them.  Here the cyclic field may be the
          most interesting because the number of the iterations in
          the cycle is undetermined.  If it were possible to reduce
          the number to $0$, then one would end up with a picture
          like Fig. \ref{RCST1} except that the $R$ and $C$ base of
          the frame would be identified in some sense with the gauge
          invariant representations described in the frame.  Whether
          this is possible, or even desirable, or not, is a problem left to
          future work.

          Another approach to connecting the frame field to physics
          is based on noting that the independence of  physical
          theories from the properties of different real and complex
          number representations can be expressed using notions of
          symmetry and invariance.  This is that
          $$\begin{array}{l}\mbox{\emph{All physical theories to date
          are invariant}} \\\mbox{\emph{under all $SU(2)$ gauge
          transformations of the qubit based}} \\ \mbox{\emph{representations
          of real and complex numbers.}}\end{array}$$ Note that the
          gauge transformations  apply not only to the qubit string
          states but also to the arithmetic relations and operations
          on the string states, Eqs. \ref{=AU} and \ref{addAU}, to sequences
          of states, to the Cauchy condition Eq. \ref{cauchyU}, and
          to the definition of the basic field operations on the real
          numbers.

          The symmetry of physical theories under these gauge
          transformations suggests that it may be useful to
          drop the invariance and consider at the outset candidate
          theories that break this symmetry.  These would be
          theories in which some basic aspect of physics is
          representation dependent.  One approach might be to look
          for some type of action whose minimization, coupled with
          the above requirement of gauge invariance,  leads to some
          restriction on candidate theories.   This gauge theory
          approach is yet another aspect to investigate in the
          future.

          \section{Discussion}

          There are some points of the material presented here that
          should be noted.  The gauge transformations described here
          apply to finite strings of qubits and their states.
          These are the basic objects.  Since these can be used to
          represent natural numbers, integers, and rational numbers
          in quantum mechanics, one can, for each type of number,
          describe different representations related by $SU(2)$ gauge
          transformations on the qubit string states. Here this
          description was extended to sequences of qubit string
          states that satisfied the Cauchy condition to give
          different representations of the real numbers.

          A reference frame was associated to each real and complex
          number representation.  Each frame contains a
          representation of all physical theories as mathematical
          structures based on the real and complex number
          representation base of the frame. If the space time manifold is
          considered to be a $4-tuple$ of the real numbers, then each
          frame includes a representation of space time as a
          $4-tuple$ of the real number representation.

          If one takes this view regarding space time, it follows
          that  for all frames with an ancestor frame, an observer
          outside the frame field or an observer in an ancestor
          frame sees that the points of space time in each descendent
          frame have structure as each point is an equivalence class of
          Cauchy sequences of (or a Cauchy operator on) states of qubit
          strings.  It is somewhat disconcerting to regard space
          time points as having structure.  However this structure
          is seen only by the observers noted above. An observer in
          a frame $F$ does not see his or her own space time points
          as having structure because the real numbers that are
          the base of his frame do not have structure. Relative
          to an observer in frame $F$, the space time base of the frame
          is a manifold of featureless points.

          It should also be noted that even if one takes the view
          that the space time manifold is some noncompact, smooth
          manifold that is independent of $R^{4},$ one still has the
          fact that functions from the manifold to the real numbers
          are frame dependent in that the range set of the
          functions is the real number representation base of the
          frame. Space time metrics are good examples of this.  As is
          well known they play an essential role in physics.

          In quite general terms, this work is motivated by the need
          to develop a coherent theory that treats mathematics and physics
          together as a coherent whole \cite{BenTCTPM}. It
          may be hoped that the approach taken here that describes
          fields of frames based on different representations of
          real and complex numbers will shed light on such a theory.
          The point that these representations are based on
          different representations of states of qubit strings shows
          the basic importance of these strings to this endeavor.

          Finally it should be noted that the  structure of
          frames emanating from  frames has nothing to do with the
          Everett Wheeler view of multiple universes \cite{Everett}.
          If these multiple universes exist, then they would exist
          within each frame in the field.

          \section*{Acknowledgements}
          This work was supported by the U.S. Department of Energy,
          Office of Nuclear Physics, under Contract No. W-31-109-ENG-38.

           \end{document}